\documentclass[aps,pra,reprint,superscriptaddress,noshowpacs]{revtex4-1}

\usepackage{graphicx}
\usepackage{amsmath}
\usepackage{mathtools}
\usepackage{textpos}

\DeclareMathOperator{\Tr}{Tr}

\begin{document}
\title{Frequency-encoded photonic qubits for scalable quantum information processing}

\author{Joseph M. Lukens}
\email{lukensjm@ornl.gov}
\author{Pavel Lougovski}
\email{lougovskip@ornl.gov}
\affiliation{Quantum Information Science Group, Oak Ridge National Laboratory, Oak Ridge, Tennessee 37831, USA}

\date{\today}

\begin{abstract}
Among the objectives toward large-scale quantum computation is the quantum interconnect: a device which uses photons to interface qubits that otherwise could not interact. However, current approaches require photons indistinguishable in frequency---a major challenge for systems experiencing different local environments or of different physical compositions altogether. Here we develop an entirely new platform which actually exploits such frequency mismatch for processing quantum information. Labeled ``spectral linear optical quantum computation'' (spectral LOQC), our protocol offers favorable linear scaling of optical resources and enjoys an unprecedented degree of parallelism, as an arbitrary $N$-qubit quantum gate may be performed in parallel on multiple $N$-qubit sets in the same linear optical device. Not only does spectral LOQC offer new potential for optical interconnects; it also brings the ubiquitous technology of high-speed fiber optics to bear on photonic quantum information, making wavelength-configurable and robust optical quantum systems within reach.
\end{abstract}

\maketitle

\section{Introduction}
\label{sec1}
The basic building block of the long-sought quantum computer is the quantum bit, or qubit: a two-dimensional quantum system which can be controlled, read, and entangled with other qubits \cite{Nielsen2000}. In light of the myriad qubit platforms currently in contention \cite{Ladd2010}, it is likely that the ultimate quantum computer will combine several technologies as a hybrid system in which dissimilar and physically separated qubits are joined via universal quantum interconnects---a network dubbed the ``quantum internet'' \cite{Kimble2008, Pirandola2016}. In this vision, quantum information must be transmitted over appreciable distances, a task for which photons prove uniquely capable. Photonic qubits operate equally well at cryogenic and room temperatures, suffer from virtually no decoherence, and are readily manipulated with optical components. Add to these the fact that they travel at the speed of light, and photons provide unrivaled performance in interfacing separate quantum systems.

The ideal photonic interconnect should also be capable of full-fledged quantum information processing in the optical domain, facilitating complex operations between heterogeneous material qubits which---whether because of thermal, spectral, or spatial incompatibility---cannot interact directly. However, the advantages which make photons such sound quantum information carriers come at a cost: forcing two single photons to interact, as required for multiqubit operations, is notoriously difficult with standard optical nonlinearities. In 2001 Knill, Laflamme, and Milburn (KLM) \cite{Knill2001} revolutionized photonic quantum computing by showing that such nonlinearities can actually be realized through detection, deriving a universal quantum computing architecture based on phase shifters, beam splitters, and  photon counters---so-called ``linear optical quantum computation'' (LOQC). Yet the spatial encoding of conventional LOQC proves entirely ill-suited for interconnecting heterogeneous qubits; not only is it unavailable in single-mode optical fiber, the most robust and efficient medium for long-distance optical transmission, but the use of a single frequency (or at least a single spectral mode) precludes direct coupling into and out of materials possessing distinct resonances.

In this paper, we propose and theoretically develop a photonic computing platform which addresses both problems. Instead of viewing frequency mismatch as an \emph{obstacle} to optical interconnects, we deem it an \emph{opportunity} and capitalize on it for encoding quantum information. Termed ``spectral LOQC,'' our protocol consists of photonic qubits which occupy two discrete spectral modes and are operated on by Fourier-transform pulse shapers and electro-optic phase modulators, standard components in classical telecommunications with purely electrical control parameters. Since each qubit shares a common spatial mode, stability against environmental fluctuations is assured, and all photons can be transmitted over long distances in well-established fiber-optic networks. This signifies a major departure from spatial/polarization-mode LOQC, where stabilizing relative phase between separate spatial modes is crucial and requires a non-trivial amount of work.  And unlike previous time-frequency computing proposals \cite{Humphreys2013,Brecht2015}, which rely on ultrafast pulsed modes, our use of narrowband modes reduces speed requirements in both manipulation and detection, and can interface directly with disparate atomic levels. Our approach thereby offers new opportunities for both interfacing matter qubits and strengthening time-frequency quantum information in general.

\section{Protocol Components}
Due to their large Hilbert spaces and compatibility with classical telecommunicatons, the time and frequency characteristics of single photons have garnered increasing interest as resources in quantum information processing: chronocyclic protocols for quantum cryptography have exploded in recent years, with advances in both theory \cite{Qi2006, Nunn2013a, Mower2013} and experiment \cite{AliKhan2007, Lee2014b, Zhong2015}. However, it is one thing to transmit and measure spectro-temporal modes, as required in quantum \emph{key distribution}; it is an entirely different matter to mix and manipulate these modes at the single-photon level, mandatory for quantum \emph{computation}. Clever protocols appear possible, though: quantum pulse gates have been proposed for chronocyclic operations \cite{Brecht2015}, and a complete time-mode-based LOQC proposal makes use of polarization rotation and delay to mix optical time bins located in a single spatial mode \cite{Humphreys2013}. Yet because of the ultrafast nature of the modes involved in either example, sophisticated nonlinear optical control is required, either by specification \cite{Brecht2015}, or by necessity \cite{Humphreys2013} due to the difficulty in realizing sufficiently long polarization-induced delays.

To remove such restrictions, here we consider monochromatic frequency bins, such as those in classical dense wavelength-division multiplexing. Spectral modes offer additional benefits absent in temporal modes, such as straightforward high-resolution measurement, reduced constraints on detector jitter, and the prospect for temporally simultaneous processing of multiple modes. For example, time-to-frequency conversion has been used to extract temporal information too fast for direct detection \cite{Donohue2013, Donohue2014}, highlighting a practical advantage of working in frequency space instead of time. But while spectral quantum modes have been considered in the context of continuous-variable cluster-state quantum computing \cite{Pysher2011,Humphreys2014,Roslund2014}---an alternative approach to the standard circuit model---no true spectral LOQC protocol has been formalized.

\begin{figure}[t]
\includegraphics[width=3in]{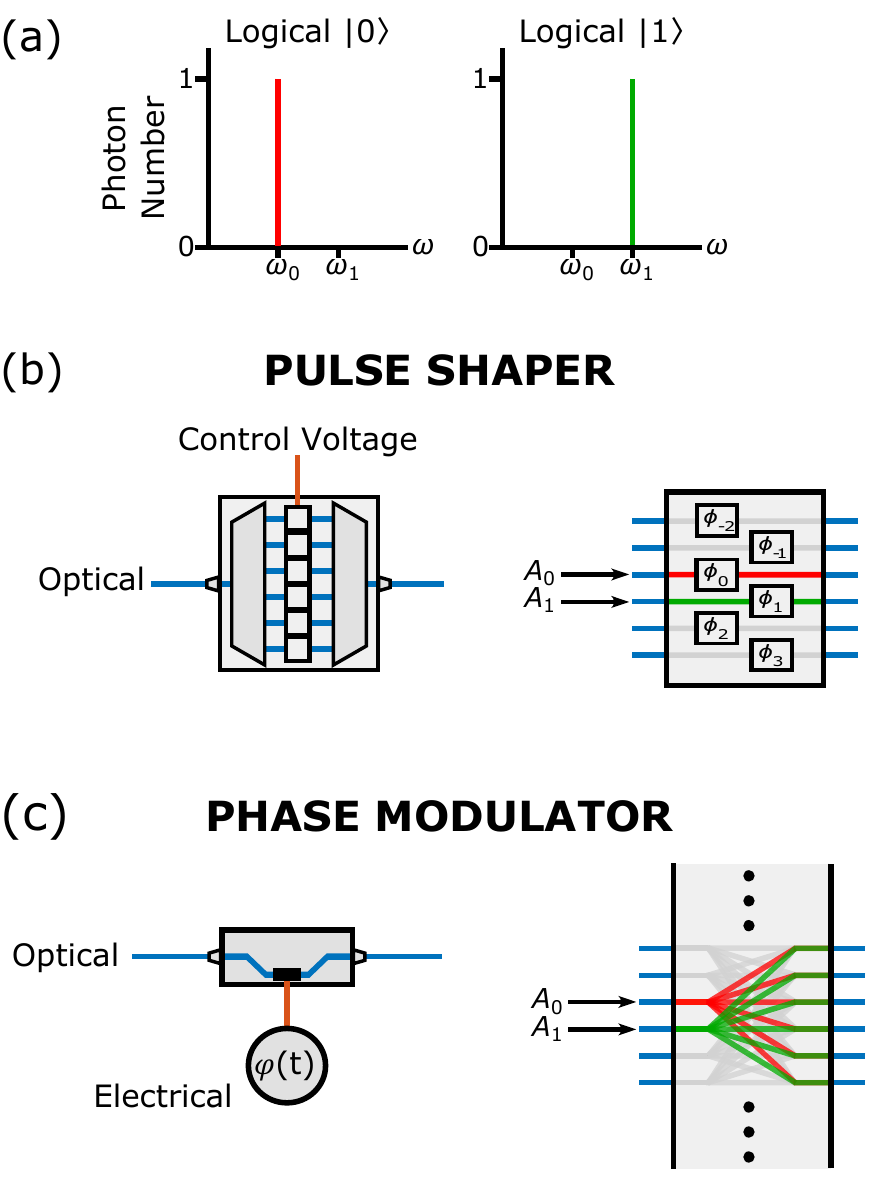}
\caption{\label{fig1} Building blocks for spectral LOQC. (a) Dual-rail qubit encoding. A single photon corresponds to $|0\rangle_L$ or $|1\rangle_L$ depending on which one of two modes it occupies. (b) Fourier-transform pulse shaper. This applies arbitrary phases to each spectral mode, physically by separating and recombining frequency components (left) and conceptually as a multimode element operating on all rails individually (right). (c) Electro-optic phase modulator. This device (left) applies an arbitrary temporal phase periodic at the inverse mode spacing . In rail form, the modulator acts as mode mixer which can move photons across frequency states (right). The labels $A_0$ and $A_1$ mark the zero and one modes for a representative qubit.}
\end{figure}

The Hilbert space in which we work consists of the countably infinite set of modes with frequency $\omega_n$, with $n$ any integer. Spatial and polarization degrees of freedom are assumed constant over all frequencies and are neglected. The fixed spacing between modes is $\Delta\omega=\omega_{n+1}-\omega_n$, so that a single dual-rail qubit spanning modes $p$ and $q$ assumes the form
\begin{equation}
\label{e1}
|\psi\rangle = \alpha |0\rangle_L + \beta|1\rangle_L = \alpha |1_p 0_q\rangle + \beta|0_p 1_q\rangle,
\end{equation}
where $|\alpha|^2+|\beta|^2=1$. Here the subscript $L$ denotes the logical state, whereas $|N_p\rangle = (N!)^{-1/2} \left(\hat{a}_p^\dagger\right)^N|0_p\rangle$ corresponds to the physical Fock state with $N$ quanta occupying mode $p$. As in the original KLM proposal, the modes $p$ and $q$ need not be adjacent on the grid, but can be chosen for convenience in realizing a particular operation. Figure \ref{fig1}(a) provides a visualization of this spectral encoding; in this case, a single photon at frequency $\omega_0$ represents $|0\rangle_L$, whereas a photon at $\omega_1$ signifies $|1\rangle_L$. The computational operations we implement are best represented by their action on the positive-frequency electric-field operator \cite{Glauber1963-1,Mandel1995}, expressed before a given operation as
\begin{equation}
\label{e2}
\hat{E}_\mathrm{in}^{(+)}(t) = \sum_{n=-\infty}^{\infty} \hat{a}_n \, e^{-i\omega_n t},
\end{equation}
and after as
\begin{equation}
\label{e2-1}
\hat{E}_\mathrm{out}^{(+)}(t) = \sum_{n=-\infty}^{\infty} \hat{b}_n \, e^{-i\omega_n t},
\end{equation}
where we have neglected unimportant scaling factors, and used the symbol $\hat{a}_n$ ($\hat{b}_n$) to denote mode $n$'s annihilation operator before (after) the optical transformation.

Motivated by feasibility, we build our computational operations on components common in telecommunication networks. In general, all computing operations can be decomposed into phase shifts and cross-modal couplings. For the former, a Fourier-transform pulse shaper can be used \cite{Weiner2000, Weiner2011}, a simplified schematic of which is presented in Fig. \ref{fig1}(b). This device separates frequency modes of the input with, e.g., a diffraction grating, prism, or arrayed waveguide grating. Then the phase of each mode is manipulated by adjusting the voltage on separate pixels of a spatial light modulator (or any other phase-control element). Recombining the spectrum then leaves an arbitrary output field. From the perspective of optical quantum computing, the ideal pulse shaper applies an arbitrary phase to each spectral mode in question---essentially it can implement any diagonal unitary matrix in the frequency mode space. Explicitly, a pulse shaper transforms the electric field in Eq. (\ref{e2}) as
\begin{equation}
\label{e3}
\hat{E}_\mathrm{out}^{(+)}(t) = \sum_{n=-\infty}^{\infty}e^{i\phi_n} \hat{a}_n \, e^{-i\omega_n t},
\end{equation}
so that on a mode-by mode basis,
\begin{equation}
\label{e4}
\hat{b}_n=e^{i\phi_n} \hat{a}_n.
\end{equation}
Pulse shaping has been applied to single and entangled photons \cite{Peer2005, Dayan2007a, Zaeh2008, Poem2012, Bernhard2013, Bessire2014}, including works directly in the telecom band \cite{Lukens2013b, Lukens2013c, Lukens2014a, Lukens2014b, Odele2015}.

For the second operation, mode mixing, we make use of the electro-optic phase modulator (EOM) \cite{Wooten2000}. Based on the electro-optic effect, a voltage applied across an EOM modifies the refractive index of the material and hence the phase accrued by an optical field traveling through it. If this voltage itself oscillates at a particular frequency (typically in the range DC--100 GHz), it generates equispaced sidebands, an effect exploited to realize tunable optical frequency combs \cite{Jiang2007,Metcalf2013}. Figure \ref{fig1}(c) provides a schematic of such a device. By controlling the voltage pattern $\varphi(t)$ applied to the device, assumed periodic with frequency equal to the mode spacing, it is possible to modify the coupling between frequency slots. Mathematically speaking, if the phase modulation operation is expressed as the Fourier series $e^{i\varphi(t)} = \sum_k c_k e^{-ik\Delta\omega t}$, then the electric field transforms as
\begin{equation}
\label{e5}
\hat{E}_\mathrm{out}^{(+)}(t) = e^{i\varphi(t)} \hat{E}_\mathrm{in}^{(+)}(t) = \sum_{k=-\infty}^{\infty} \sum_{n=-\infty}^{\infty} c_k \hat{a}_n \, e^{-i(\omega_n + k\Delta\omega)t},
\end{equation}
from which the modal transformation is found to be
\begin{equation}
\label{e6}
\hat{b}_n = \sum_{k=-\infty}^\infty c_{n-k}\hat{a}_k,
\end{equation}
which is unitary on the countably infinite set of spectral modes ($\sum_k c_{m-k}^* c_{n-k} = \delta_{mn}$). A functional block representation of the EOM follows in Fig. \ref{fig1}(c) (right), where each rail denotes a separate frequency mode. In this example, the single-qubit amplitudes in modes $A_0$ and $A_1$ interfere analogously to a spatial interferometer. As with the pulse shaper, experiments over the last decade \cite{Kolchin2008, Harris2008, Sensarn2009a, Belthangady2009, Belthangady2010, Olislager2010, Karpinski2016, Wright2016} have confirmed that nonclassical single-photon states do respond to electro-optic modulation as expected from this theory. 

However, because the coupling is effected on all modes globally, rather than in pairs, the nature of this mixing is markedly different than that of the beamsplitters used to argue scalability in spatial LOQC \cite{Reck1994}. Nevertheless, a sequence of alternating pulse shapers and EOMs is sufficient to reproduce any matrix transformation. As demonstrated by a recent constructive proof \cite{Huhtanen2015}, any $N\times N$ complex matrix can be factored exactly as the product of no more than $2N-1$ circulant and diagonal matrices---or equivalently of $2N-1$ diagonal matrices spaced by DFT matrices. As described in the following section, such a decomposition is precisely that provided by EOMs and pulse shapers when discretized for numerical simulation, thereby implying that the number of components needed to implement an arbitrary unitary transformation on $N$ spectral modes scales like $\mathcal{O}(N)$. In contrast, the scaling of optical components for spatial- or polarization-encoded LOQC is quadratic [$\mathcal{O}(N^2)$]. Such an improvement in our protocol can be understood intuitively through complexity arguments \cite{Miller2013}. In general, a mode transformation between $N$ input and $N$ output modes requires $\mathcal{O}(N^2)$ free parameters. And whereas the transformations of spatial beamsplitters and phase shifters are fixed by only one or two numbers, those provided by pulse shapers and EOMs offer $\mathcal{O}(N)$ independent real parameters. Thus one can view the linear scaling as a consequence of the fact that the building blocks of spectral LOQC operate on all modes simultaneously, rather than fixed subsets.

Finally, it is important to note that spectral mode mixing could alternately be accomplished through interactions mediated by a classical field in a nonlinear medium, as in a recent experiment exploiting Bragg scattering to transform a single spectral qubit like ours \cite{Clemmen2016}. However, EOMs are simpler and more scalable, for they require only a single electrical control, produce no noise photons from powerful optical fields, and are linear in the sense that their characteristics are independent of the number of optical photons passing through them \cite{Miller2013b}. Similar considerations apply also to acousto-optic modulators, which are mathematically equivalent to two-mode beam splitters \cite{Jones2006}; but because of their noncollinear interaction geometry, such modulators do not support single-spatial-mode operation. Accordingly, of these options, only EOMs satisfy both linearity and spatial purity, and so we classify our computing protocol built on pulse shapers and EOMs as truly a form of LOQC.

\section{Deriving Unverisal Gate Set}
\label{sec4}
In the spirit of spatial KLM computing, we assume in our protocol ancillary single photons and perfect photon-number-resolving detectors; such additional resources are required to nondestructively mark completion of a two-qubit gate. We also allow for vacuum modes which can be populated during the gate operation, later showing that the number of such modes can be restricted in a practical gate. Pulse shapers and EOMs are modeled as matrices acting on our spectral mode space, truncated to $M$ modes; if $M$ is sufficiently large, this well approximates the transformation on the interior computational modes. Specifically, each pulse shaper acts as a diagonal matrix $D$ consisting of complex-exponential elements, and each EOM acts as a unitary diagonal matrix $\tilde{D}$ in time, transformed to frequency by the DFT matrix $F$ whose elements are defined via $F_{nk}= M^{-1/2} e^{2\pi i nk/M}$. Thus the spectral transformation realized by an EOM is given by $F \tilde{D} F^\dagger$, where the $M$ phases in $\tilde{D}$ represent samples in a single temporal period---an approximation which holds when the number of samples is sufficient to model the phase smoothly.

We apply pulse shapers and modulators in an alternating sequence, and focus on how they transform an input state in the computational subspace of the Hilbert space $\mathcal{H}_{N}^{n}$ for $n$ photons in $N$ modes of interest. A series of $R$ pulse shapers and $R$ EOMs produces the following $M\times M$ unitary mode transformation acting on the $N$ computational and $M-N$ ancilla modes' annihilation operators:
\begin{equation}
\label{e7}
V = F \tilde{D}_R  F^\dagger D_R \cdot\cdot\cdot F \tilde{D}_2  F^\dagger D_2 F \tilde{D}_1  F^\dagger D_1 .
\end{equation}

With this mode transformation in hand, our procedure for analyzing gate performance follows that of \cite{Uskov2009}; see Appendix \ref{app1} for details. From the mode operator we derive the equivalent state transformation $W$, represented in the Fock basis and projected onto the detection of any ancillas. Then we compare it to the target state transformation $T$ via the fidelity $\mathcal{F}$, defined according to the Hilbert-Schmidt inner product:
\begin{equation}
\label{e8}
\mathcal{F} = \frac{\Tr(W^\dagger T) \Tr(T^\dagger W) }{\Tr(W^\dagger W) \Tr(T^\dagger T) }.
\end{equation}
This is equal to unity when $W$ is proportional to $T$. The success probability $\mathcal{P}$ is then defined as
\begin{equation}
\label{e9}
\mathcal{P} = \frac{\Tr(W^\dagger W) }{ \Tr(T^\dagger T) },
\end{equation}
which is independent of the input state as $\mathcal{F}\rightarrow 1$. So, after specifying the mode positions of all qubits, ancillas, and detectors, and choosing the number of elements for a given operation, we run a numerical optimization routine searching for the $2RM$ phases that maximize the success probability $\mathcal{P}$ while preserving fidelity $\mathcal{F}=1$.

\begin{figure*}
\includegraphics[width=7in]{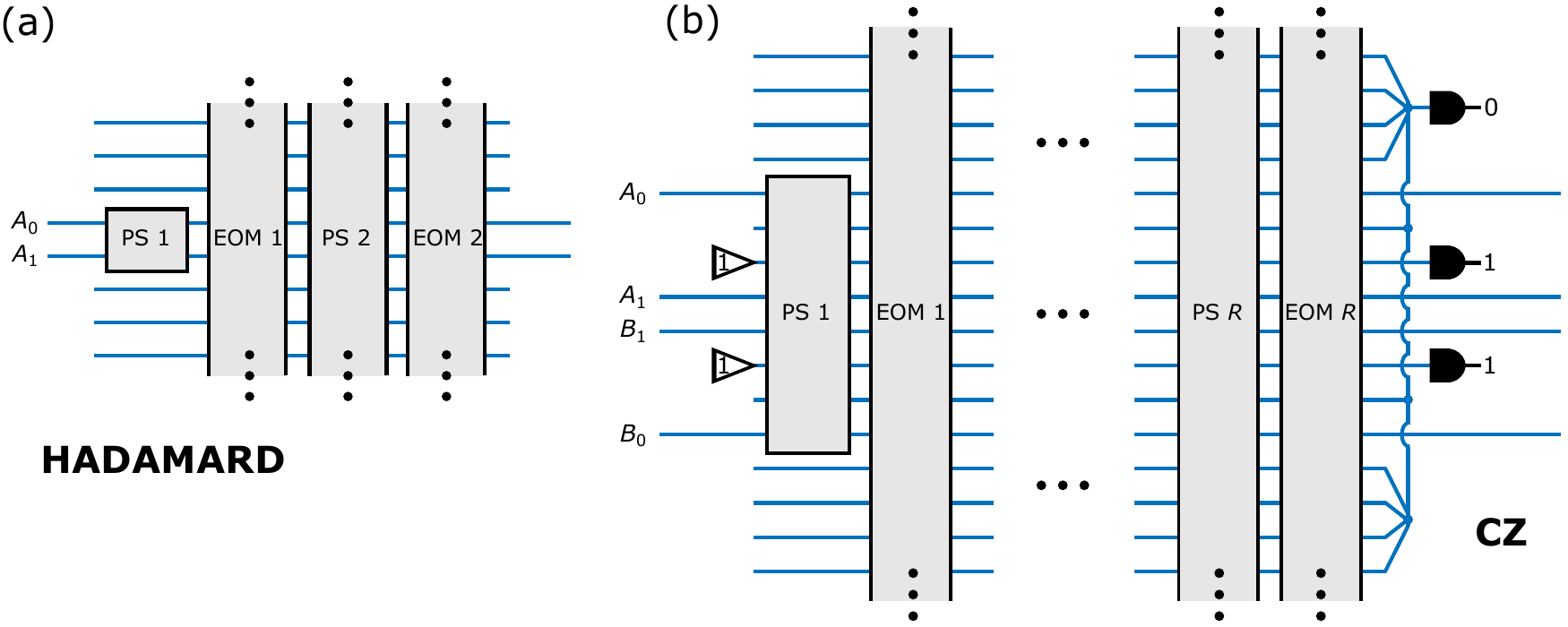}
\caption{\label{fig2} Schematics of spectral LOQC gates. Each rail represents a distinct frequency mode, in increasing value from top to bottom. The logical zero and one modes for the first qubit are labeled $A_0$ and $A_1$; those for the second qubit are $B_0$ and $B_1$. The labels PS and EOM denote pulse shaper and electro-optic phase modulator, respectively. (a) Hadamard gate. With two pulse shapers and two EOMs, this operation succeeds with probability 1, requiring no ancillas. (b) \textsc{cz} gate. Two ancilla photons are loaded in the modes adjacent to $A_1$ and $B_1$, and all photons propagate through a series of $R$ pulse shaper/EOM pairs. The spectrally resolved detection pattern shown here then heralds successful completion of the gate.} 
\end{figure*}

Our primary order of business is to derive a universal gate set \cite{Nielsen2000}. For our purposes, the family comprising the single-qubit phase and Hadamard ($H$) operations and the two-qubit controlled-$Z$ (\textsc{cz}) gate proves most convenient. In the single-qubit case, spectral LOQC delivers the phase gate directly: a single line-by-line pulse shaper can effect arbitrary phase shifts between spectral modes, ideally with no loss. Yet the Hadamard operation is more difficult. In the logical qubit space $(|0\rangle_L, |1\rangle_L)$ it is defined: 
\begin{equation}
\label{e10}
H = \frac{1}{\sqrt{2}}
\begin{bmatrix}
    1 & 1  \\
    1 & -1
  \end{bmatrix}.
\end{equation}
The nature of electro-optic phase modulation [Eq. (\ref{e6})] makes a lossless mixer such as $H$ impossible with a single EOM; however, we do find it attainable with \emph{two}. Figure \ref{fig2}(a) shows the block form of our Hadamard gate acting on adjacent spectral rails, utilizing two pulse shapers and two EOMs. The dots above and below the latter three devices indicate potential coupling to modes beyond those in the figure. The numerical solution we obtain (plotted in Appendix \ref{app2}) possesses $\mathcal{F}=1$ and $\mathcal{P}=1$, so the gate is entirely deterministic; the qubit exits in the same modes it entered with no need for heralding detectors. The second two rows in Table \ref{t1} summarize performance of these single-qubit gates, listing the physical resources (number of pulse shapers, number of EOMs, and number of ancilla photons) as well as the key performance metrics (fidelity, probability of success, and approximate number of spectral modes required).

\begin{table}[b]
\caption{\label{t1} Summary of universal gate set for spectral LOQC.}
\begin{tabular}{c|cccc} \hline
 & Shapers- & Fidelity & Prob. & Eff. \\
Gate & EOMs-Ancillas & ($\mathcal{F}$) & ($\mathcal{P}$) & \# modes \\ \hline
Phase & 1-0-0 & 1.0000 & 1.0000 & 2 \\
$H$ & 2-2-0 & 1.0000 & 1.0000 & 8 \\ \hline
 & 2-2-2 & 1.0000  & 0.0207 & 48 \\
\sc{cz} & 3-3-2 & 0.9999 & 0.0672 & 24 \\
 & 4-4-2 & 0.9999  & 0.0735 & 18 \\ \hline
\end{tabular}
\end{table}

To complete the universality proof, we construct a two-qubit \textsc{cz} gate, alternatively called controlled-phase (\textsc{cphase}) or controlled-sign (\textsc{cs}). Not only are the properties of the \textsc{cz} gate well studied in LOQC, providing a benchmark for clear comparison, but the \textsc{cz} gate is also symmetric in both photons---an aid in our design procedure. In the logical basis of spectral qubits $A$ and $B$, ($|0_A 0_B\rangle_L, |0_A 1_B\rangle_L, |1_A 0_B\rangle_L, |1_A 1_B\rangle_L$), the matrix representing this operation is
\begin{equation}
\label{e11}
U_\textsc{cz} = \begin{bmatrix}
    1 & 0 & 0 & 0  \\
    0 & 1 & 0 & 0 \\
    0 & 0 & 1 & 0 \\
    0 & 0 & 0 & -1 
  \end{bmatrix}.
\end{equation}
The most efficient nondestructive \textsc{cz} gate in KLM computing known to date, making use of two ancillary photons, was introduced by Knill in 2002: it succeeds with probability $\mathcal{P}=2/27\approx 0.0741$ \cite{Knill2002}. Although better implementations are not excluded theoretically, extensive numerical searches indicate that this is indeed the optimal gate with nonentangled ancillas \cite{Uskov2009}. Since the specific form of our linear optical operation [Eq. (\ref{e7})] is subsumed within the general search space considered in \cite{Uskov2009}, we anticipate being able to do no better than  $\mathcal{P}=7.41\%$. The goal, then, is to come as close as possible to this limit with minimal resources.

A schematic of our proposed \textsc{cz} gate is provided in Fig. \ref{fig2}(b). The logical $|0\rangle_L$ and $|1\rangle_L$ modes for each photon are separated by two auxiliary spectral modes, one of which is loaded with an ancilla photon (triangle). The photons then propagate through a series of $R$ pulse-shaper/modulator pairs, followed by detection in the ancilla modes; the specific coincidence pattern in Fig. \ref{fig2}(b) (zero photons in the top detector, one photon in the middle, one photon on the bottom) signifies successful gate operation. Because of its probabilistic nature, we do need to look for photons in the adjacent vacuum modes, but only to ensure that no photons have escaped the computational space. Thus we can employ a single spectral ``bucket'' detector [top detector of Fig. \ref{fig2}(b)] which checks whether any photon leaks into the initially unoccupied modes. The practical number of modes to which this detector must respond is determined by how strongly coupled they are to the six input modes loaded with photons. To express it another way, the detection pattern in Fig. \ref{fig2}(b) guarantees that two photons exit in the four computational modes, and the spectro-temporal modulation patterns then must ensure that when this happens, the two have undergone a \textsc{cz} operation.

Starting with $R=2$, we constrain the fidelity to $\mathcal{F}\geq0.9999$ (comfortably within the fault-tolerant limit for LOQC under any definition~\cite{Devitt2013}) and numerically determine the phases required to maximize the success probability $\mathcal{P}$. The results for our optimization as $R$ is increased are presented in the last three rows of Table \ref{t1}. At $R=4$, the success probability is within 1\% of the KLM limit, indicating optimal performance. And even at $R=3$, the success probability surpasses the slightly less optimal gate from the original KLM proposal, which had $\mathcal{P}=1/16=0.0625$ \cite{Knill2001}. (Appendix \ref{app2} plots the specific EOM and pulse shaper phases for each solution.) With parameters for a two-qubit spectral \textsc{cz} gate now established, the universal gate set is complete. While our spectral version does require additional vacuum modes (quantified in the next section), the success probabilities and ancilla photon requirements match the best known in general linear optics; this represents the main finding of our paper.

\section{Practical Considerations}
One of the potential advantages of spectral LOQC is its amenability to massive parallelization; that is, a gate can be applied to spectrally distinct qubits simultaneously, all within the same spatial mode and propagating through the same components. However, the fact that both single- and two-qubit operations couple to adjacent modes implies that guard bands between qubits may be required, limiting how densely such qubits can be packed. In order to address this issue explicitly, here we quantify the bandwidth needed to maintain high-performance operation. The procedure is to introduce bandpass filtering in each of our pulse shapers, incrementally increasing the bandwidth until $\mathcal{F}$ and $\mathcal{P}$ reach their asymptotically optimum values. We do not allow any photon to stray from the specified band at any point in the circuit. Thus, the smallest bandwidth at which the performance reaches its optimal value gives a conservative estimate for the required gate band.

\begin{figure}
\includegraphics[width=3.5in]{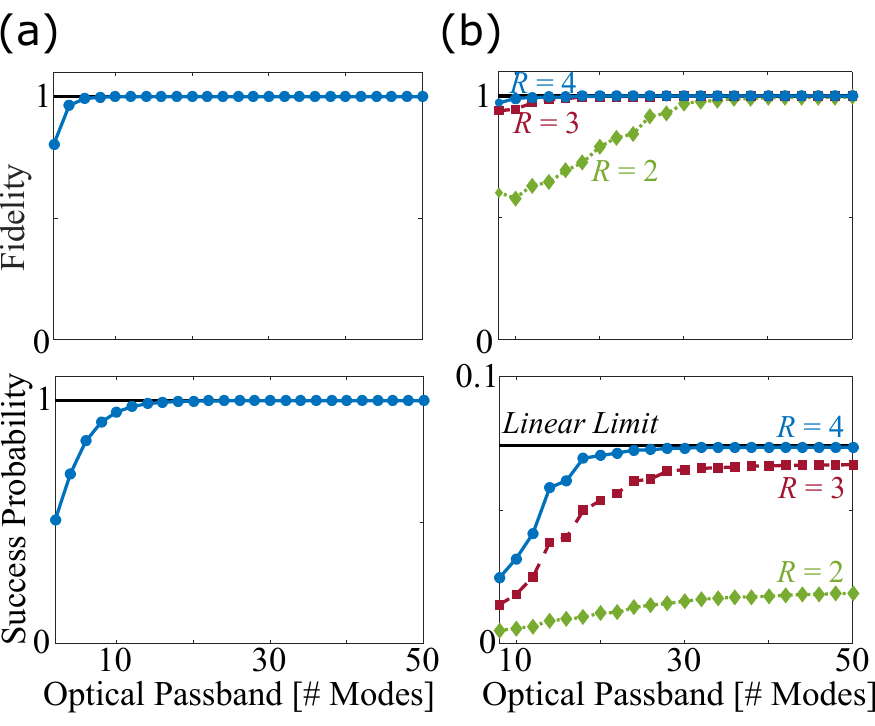}
\caption{\label{fig3} Bandwidth scaling. These plots show the performance of the spectral Hadamard and \textsc{cz} gates as a function of optical bandwidth, in terms of fidelity $\mathcal{F}$ and success probability $\mathcal{P}$. (a) Hadamard gate: fidelity (top) and success probability (bottom). (b) \textsc{cz} gate: fidelity (top) and success probability (bottom). All three configurations from Table \ref{t1} are tested: $R=2$ (green), $R=3$ (red), and $R=4$ (blue). The black line in the bottom plot marks the value 0.0741---the best linear optical \textsc{cz} gate known to date.} 
\end{figure}

The passbands required for success probabilities ${>90\%}$ of their asymptotic values are listed in the last column of Table \ref{t1}. Since the phase gate preserves photon number in each mode, nothing beyond the logical basis is required. Such is not the case for the Hadamard gate, whose full results are summarized in Fig. \ref{fig3}(a). As the gate passband is increased in multiples of two modes, the fidelity reaches 99\% for a passband of six (two vacuum modes on either size of the qubit), while the success probability attains 90\% of its asymptotic value for an eight-mode band. Thus, the optical passband need only be chosen about eight times the logical mode spacing in order to reach good performance. Our simulations on the $\textsc{cz}$ gate follow the same approach, but now we look at all three possible configurations from Table \ref{t1} ($R=2,3,4$). The results displayed in Fig. \ref{fig3}(b) reveal a fascinating dependence on the number of elements. Not only does a larger value of $R$ attain a greater asymptotic success probability ($R=4$ just reaches the linear limit marked by a black line in the bottom plot), but it also does so more efficiently in terms of the effective number of modes. The case $R=4$ needs only 18 optical modes to attain 90\% of its asymptotic success probability, compared to 24 and 48 for $R=3$ and $R=2$, respectively. This scaling proves especially interesting because one might initially conjecture that more EOMs would actually utilize larger bandwidths, for each successive element couples the input to more modes. Instead, the observed performance hints at the positive correlation between success probability and keeping photons ``close'' to the computational modes; with more parameters available in the optimization, larger values of $R$ can realize better performance all around. We therefore suspect that this tradeoff might prove more general in chronocyclic manipulations: fewer devices require more ancilla modes to maintain comparable performance. 

\begin{figure*}
\includegraphics[width=7in]{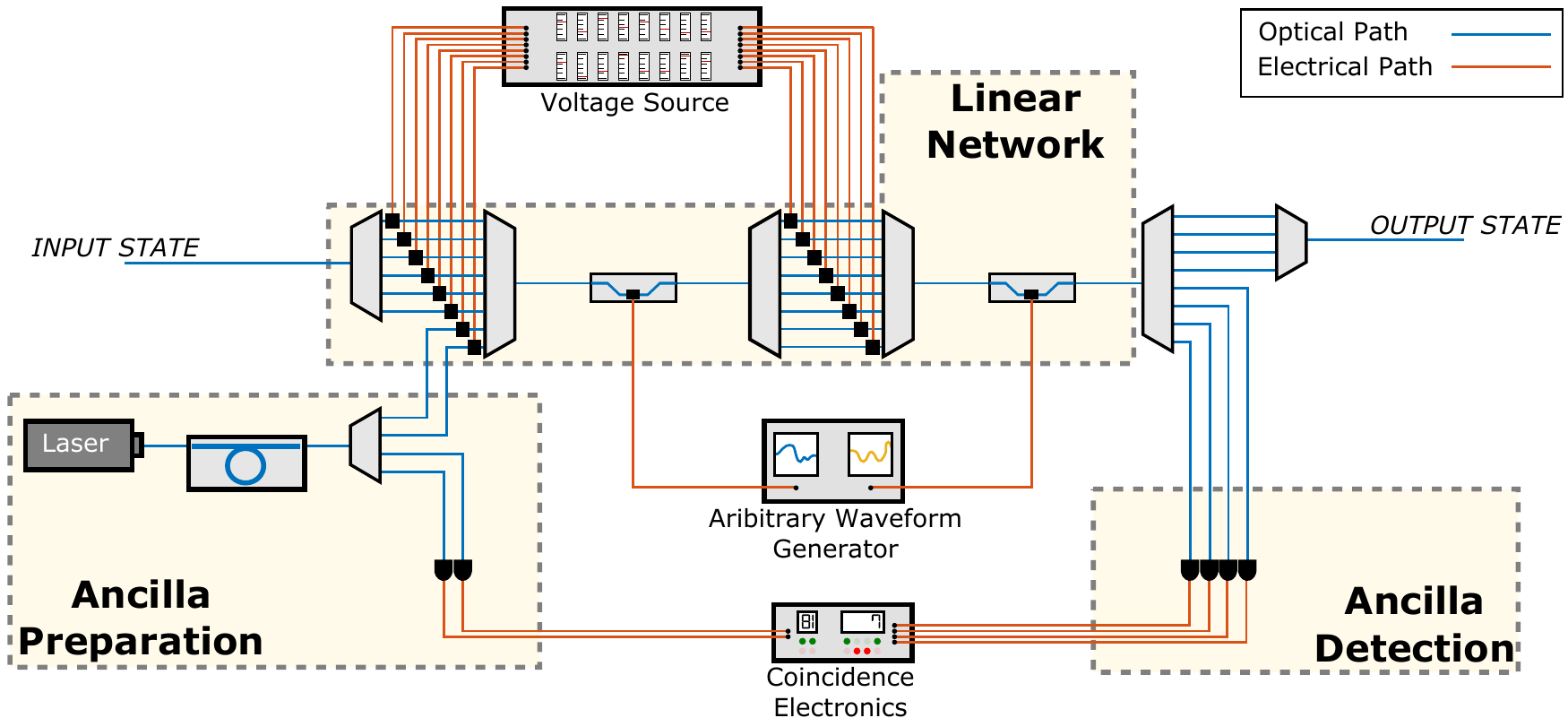}
\caption{\label{fig4} Outline of potential spectral LOQC experiment. The setup contains three basic systems, concerned with generating auxiliary photonic resources (Ancilla Preparation), manipulating the frequency modes with pulse shapers and EOMs (Linear Network), and detecting ancillas to determine successful operation (Ancilla Detection).} 
\end{figure*}

Looking toward the ultimate implementation of this new protocol, we now outline performance requirements and current practical capabilities. Figure \ref{fig4} sketches a possible realization of spectral LOQC, where we assume computational photons as input from some other source, as would be the case in a photonic interconnect. The required ancillas are generated by spontaneous four-wave mixing in a resonator \cite{Grassani2015,Reimer2016}, followed frequency-resolved detection (``Ancilla Preparation''). By choosing the spectral bins appropriately, ancillas in the desired frequency modes can be heralded. The photons are then combined via spectral multiplexers into the circuit itself (``Linear Network''), which consists of pulse shapers with voltage-controlled phases and EOMs driven by a high-speed arbitrary waveform generator. Upon exiting the computational circuit, the auxiliary modes are separated and detected (``Ancilla Detection''); if the desired coincidence pattern is attained, the operation succeeds and the output consists of transformed spectral qubits.

In order to analyze how such a circuit could perform, we focus on the elements unique to spectral LOQC (the ``Linear Network'' in Fig. \ref{fig4}), since photon preparation and detection are common to all versions of KLM quantum computing. The challenges these devices pose fall into three basic categories. (i) Spectral resolution: each pulse shaper must be able to separate frequency modes and apply arbitrary phases with minimal crosstalk. (ii) Microwave bandwidth: the electronic waveform generators---and the EOMs driven by them---must be able to reproduce the required phase modulation patterns. (iii) Loss: although phase modulation is theoretically unitary, realistic components can apply significant attenuation, particularly in waveguide coupling or diffractive elements. Perhaps predictably, these three pragmatic concerns often are at odds with each other. Widely spaced frequency modes would ease spectral resolution requirements, but tighten demands on modulation bandwidth; and pulse shapers with extremely high resolution often inflict large optical losses. 

First let us consider possibilities with available discrete components. For example, commercial fiber-pigtailed pulse shapers offer 10-GHz resolution and support roughly 500 separate channels across the optical telecommunication band around 1550 nm \cite{Finisar2015}. If we thus take 10 GHz as the mode spacing, we require an arbitrary microwave signal generator with $\sim$50-GHz bandwidth (Nyquist sampling rate of $\sim$100 GSa/s) for the $H$ and $R=4$ \textsc{cz} solutions in Table \ref{t1} \cite{Note}; a recent product comes very close, offering 92 GSa/s \cite{Keysight2016}. Finally, lithium niobate phase modulators with speeds exceeding 100 GHz are readily available \cite{EOSpace2016}, indicating that the spectro-temporal phase control required for our computing protocol is realizable with current technology. Such a finding is extremely important, for it justifies the reasonableness of spectral LOQC with purely electrical controls. Moreover, with 500 distinct channels, the effective bandwidth of our $R=4$ \textsc{cz} gate (18 modes) implies that one could pack approximately 28 \textsc{cz} gates in parallel without increasing the number of devices.

Unfortunately, the linear insertion loss of these discrete optical devices is relatively high, on the order of a few dB per element. But since a primary source of loss in fiber-pigtailed components is in coupling from and into the optical fiber mode, better performance could be possible with on-chip elements whose spatial modes are precisely matched. As an added benefit, the compactness would promote practicality by making the entire platform much more scalable. Research in both on-chip pulse shaping and electro-optic modulation has exploded in recent years, motivated primarily by classical optical communication. For example, microring filter banks \cite{Agarwal2006, Khan2010, Wang2015} and arrayed waveguide gratings (AWGs) \cite{Okamoto1999, Fontaine2007, Heck2008, Tahvili2013} have enabled on-chip line-by-line pulse shaping of optical fields. AWGs in particular seem well-suited for our application; with as many as 512 channels demonstrated in one example \cite{Cheung2014} and mode spacings down to 5 GHz in another \cite{Jiang2009}, the bandwidth and spectral resolution are competitive with high-performance bulk dispersers. While the total loss of pulse shapers based on AWGs has been high ($>$10 dB is common \cite{Okamoto1999,Heck2008,Tahvili2013}), clever AWG designs in other contexts have been able to reach insertion losses as low as $\sim$0.5 dB \cite{Watanabe2005}, so there is promise for driving down losses in such spectral shapers.

Integration of high-speed EOMs with AWGs has witnessed monumental progress as well, specifically in commercial wavelength-multiplexed transceivers \cite{Nagarajan2010, Kish2011}. Coupled with the independent research in chip-scale resonator-based photon sources \cite{Clemmen2009, Azzini2012, Grassani2015, Jiang2015,Reimer2016} and superconducting detectors \cite{Heeres2013, Najafi2015}, a fully integrated platform for spectral LOQC seems feasible in the future. The design requirements for on-chip chronocylic shaping do differ for quantum systems compared to classical ones; for example, whereas classical telecommunication designs can mobilize amplifiers to counteract loss, a quantum network cannot, making throughput a relatively higher priority than in the classical systems realized up to this point. For this reason, our quantum protocol suggests slightly different emphases in design, providing optimism for improved performance when devices are tailored to this application.

\section{Conclusions and Future Directions}
\label{sec7}
In this article, we have proposed and numerically derived a universal linear-optical quantum computing platform based on dual-rail frequency encoding. Our approach requires no optical nonlinearities, is compatible with classical wavelength-division-multiplexed networks, and utilizes no spatial interferometers. The necessary spectral and temporal manipulations appear possible with current technology. Moreover, one can exploit the programmability of each pulse shaper and EOM to transform a given physical arrangement into any one of several quantum circuits by modifying the electrical controls, offering interesting practical avenues toward reconfigurable quantum networks.

From a more conceptual angle, our design space of phase manipulation in Fourier dual bases offers new directions for quantum optical circuit synthesis. Rather than building quantum computations by combining one- and two-qubit gates, one can view the full quantum circuit as a sequence of chronocyclic phases. Accordingly, we anticipate significant resource reduction since each design is optimized for its specific computation, analogous to previous findings in LOQC for the case of photonic cluster-state production~\cite{BrowneRudolph2005,BodiyaDuan2006,Uskov2015}. Furthermore, although here we have specialized to time-frequency spaces, these design tools apply equally well to diffractive systems with lenses and masks, which are likewise based on successive Fourier transformations and phase manipulations---indicating pathways to take these ideas full-circle back to the spatial domain.

Yet the greatest impact of this computing protocol likely rests in the specific application of the quantum interconnect. Spectral qubits interface directly with frequency-disparate systems, can be transmitted long distances in optical fiber, and are controllable with telecommunication technology. In many ways, their advantages bear analogy to those of classical optical frequency combs \cite{Newbury2011}: just as an optical clock connects ultraprecise microwave frequencies (via the comb spacing) to optical wavelengths (via the absolute frequency), spectral LOQC can serve as the bridge between the microwave regime of material qubits and optical photons which are so adept at information transmission. And so it is our hope that quantum information processing with frequency-encoded photons will fill in yet one more piece of the puzzle that is practical quantum computing.

\begin{acknowledgments}
We thank W.~P. Grice, N.~A. Peters, B.~J. Smith, and A.~M. Weiner for valuable discussions, and A.~J. Metcalf for familiarizing us with the state of the art in on-chip pulse shaping. This work was performed at Oak Ridge National Laboratory, operated by UT-Battelle for the U.S. Department of Energy under contract no. DE-AC05-00OR22725. J.M.L. was supported through a Wigner Fellowship at ORNL.
\end{acknowledgments}

\appendix

\section{Optimization Procedure}
\label{app1}
As mentioned in the main text, the total $M$-mode transformation for a series of $R$ pulse shapers and $R$ electro-optic phase modulators (EOMs) assumes the form
\begin{equation}
\label{s1}
V = F \tilde{D}_R  F^\dagger D_R \cdot\cdot\cdot F \tilde{D}_2  F^\dagger D_2 F \tilde{D}_1  F^\dagger D_1,
\end{equation}
where the $D_k$ ($\tilde{D}_k$) are diagonal unitary matrices signifying spectral (temporal) phase modulation, and $F$ is the DFT matrix with elements $F_{nk}= M^{-1/2} e^{2\pi i nk/M}$. The input and output mode operators in the subspace of interest are connected by this matrix $V$ according to
\begin{equation}
\label{s2}
\hat{b}_j = \sum_{k=K}^{K+N-1} V_{jk} \hat{a}_k,
\end{equation}
where $K$ specifies the index of the smallest frequency in the $N$-mode subspace. 

After \cite{Uskov2009}, we first convert from the mode transformation [Eq. (\ref{s1})] to the Hilbert space $\mathcal{H}_{N}^{n}$ spanned by the photonic qubits. Let $d_\mathrm{in}$ and $d_\mathrm{out}$ signify the dimensionalities of the input and output computational photonic spaces. Then delineate the corresponding basis states by $\{ |m\rangle_\mathrm{in} : m=1,2,...,d_\mathrm{in} \}$ for the input and $\{ |l\rangle_\mathrm{out} : l=1,2,...,d_\mathrm{out} \}$ for the output; $m$ and $l$ are simply labels, \emph{not} photon numbers. For a single-qubit, $d_\mathrm{in} = d_\mathrm{out} = 2$, and we can formally write as basis states
\begin{equation}
\label{s3}
|m\rangle_\mathrm{in} = |1_{r_m}\rangle = \hat{a}_{r_m}^\dagger |0\rangle
\end{equation}
and
\begin{equation}
\label{s4}
|l\rangle_\mathrm{out} = |1_{p_l}\rangle = \hat{b}_{p_l}^\dagger |0\rangle,
\end{equation}
where $r_m$ and $p_l$ map each basis label to the particular mode occupied by the photon, which we are free to choose for convenience. By the preceding three equations, the actual $2\times2$ state transformation matrix $W$ then has elements
\begin{equation}
\label{s5}
W_{lm} = {}_\mathrm{out} \langle l | m \rangle_\mathrm{in}  = V_{p_l r_m}.
\end{equation}
That is, for a single photon, the state transformation is related directly to the mode transformation.  

For a two-qubit gate, however, the conversion from mode to state is more complicated; with two computational photons and additional ancillas, multiple pathways connect a given pair of input and output basis states. In the following we assume for specificity two ancillary photons, initially placed---and later detected---in modes $u$ and $v$ ($u\neq v$); then the input photonic states are
\begin{equation}
\label{s6}
|m\rangle_\mathrm{in} = |1_{r_m} 1_{s_m} 1_u 1_v \rangle = \hat{a}_{r_m}^\dagger \hat{a}_{s_m}^\dagger \hat{a}_{u}^\dagger \hat{a}_{v}^\dagger |0\rangle,
\end{equation}
and the dimensionality satisfies $d_\mathrm{in} =4$ (the Hilbert space of two qubits). The output states are similar, but now we allow for the possibility that the computational photons land in the same mode. Accordingly, the dimensionality is now $d_\mathrm{out} = 10$ (two photons in four modes), and the basis states become
\begin{equation}
\label{s7}
|l\rangle_\mathrm{out}  = |1_{p_l} 1_{q_l} 1_u 1_v \rangle = \frac{ \hat{b}_{p_l}^\dagger \hat{b}_{q_l}^\dagger \hat{b}_{u}^\dagger \hat{b}_{v}^\dagger}{\sqrt{1+\delta_{p_l q_l}}} |0\rangle,
\end{equation}
with the square root factor ensuring proper normalization when $p_l=q_l$. And so, combining Eqs. (\ref{s2}), (\ref{s6}), and (\ref{s7}), we arrive at a $10\times 4$ state transformation matrix given by
\begin{equation}
\label{s8}
W_{lm} = \frac{1}{\sqrt{1+\delta_{p_l q_l}}} \sum_{\substack{(w,x,y,z) = \\ \mathrm{perms}(r_m,s_m,u,v)}} V_{p_l w} V_{q_l x} V_{u y} V_{v z},
\end{equation}
where the sum extends over all 24 permutations of the four input photon modes---this accounts for every possible physical path taking four photons in modes $(r_m,s_m,u,v)$ to four in $(p_l,q_l,u,v)$. Note that the restriction to particular subspaces in Eqs. (\ref{s5}) and (\ref{s8}) implies a projection onto the vacuum in all other modes. For a probabilistic gate, one must therefore look for photons in these extra modes to ensure no qubit is lost; for a deterministic gate, no such steps are required.

\begin{figure}[t]
\includegraphics[width=3.5in]{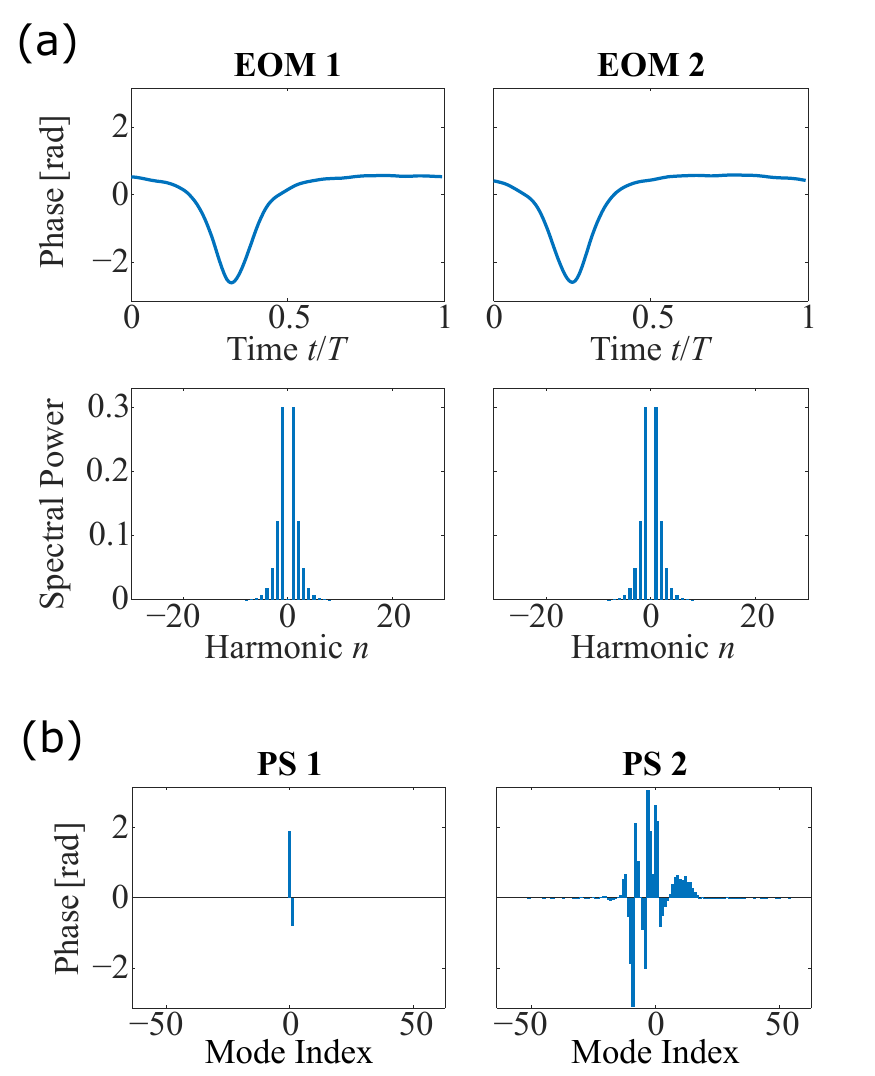}
\caption{\label{figS1} Phases applied by the components in Fig. 2(a) to realize a spectral Hadamard gate. (a) Temporal phases plotted over a single period (top) and the corresponding microwave power spectra (bottom) for the two phase modulators. (b) Pulse shaper phases, where index 0 denotes rail $A_0$ in Fig. 2(a) of the main text.} 
\end{figure}

\begin{figure}[t]
\includegraphics[width=3.5in]{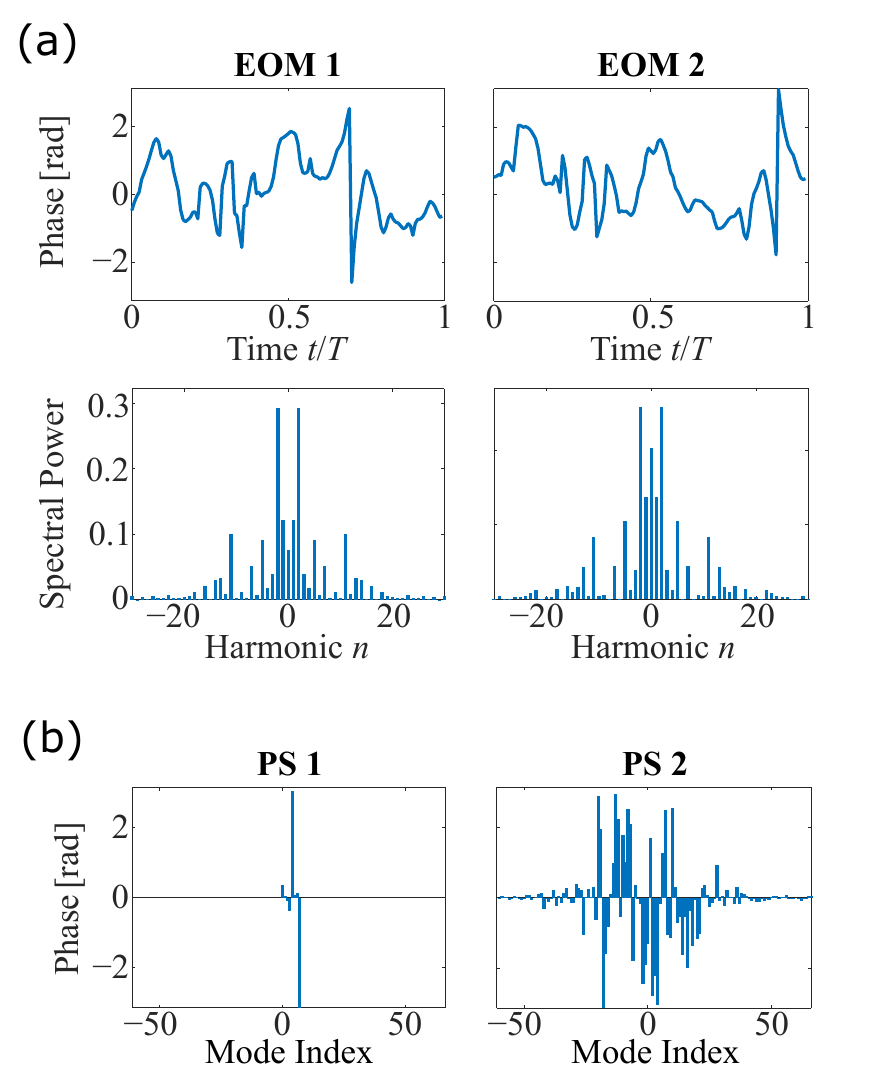}
\caption{\label{figS2} Phases applied to realize a spectral \textsc{cz} gate with $R=2$. (a) Temporal phases plotted over a single period (top) and the corresponding microwave power spectra (bottom) for the two EOMs. (b) Pulse shaper phases, where index 0 corresponds to rail $A_0$ in Fig. 2(b).} 
\end{figure}

\begin{figure*}
\includegraphics[width=6.5in]{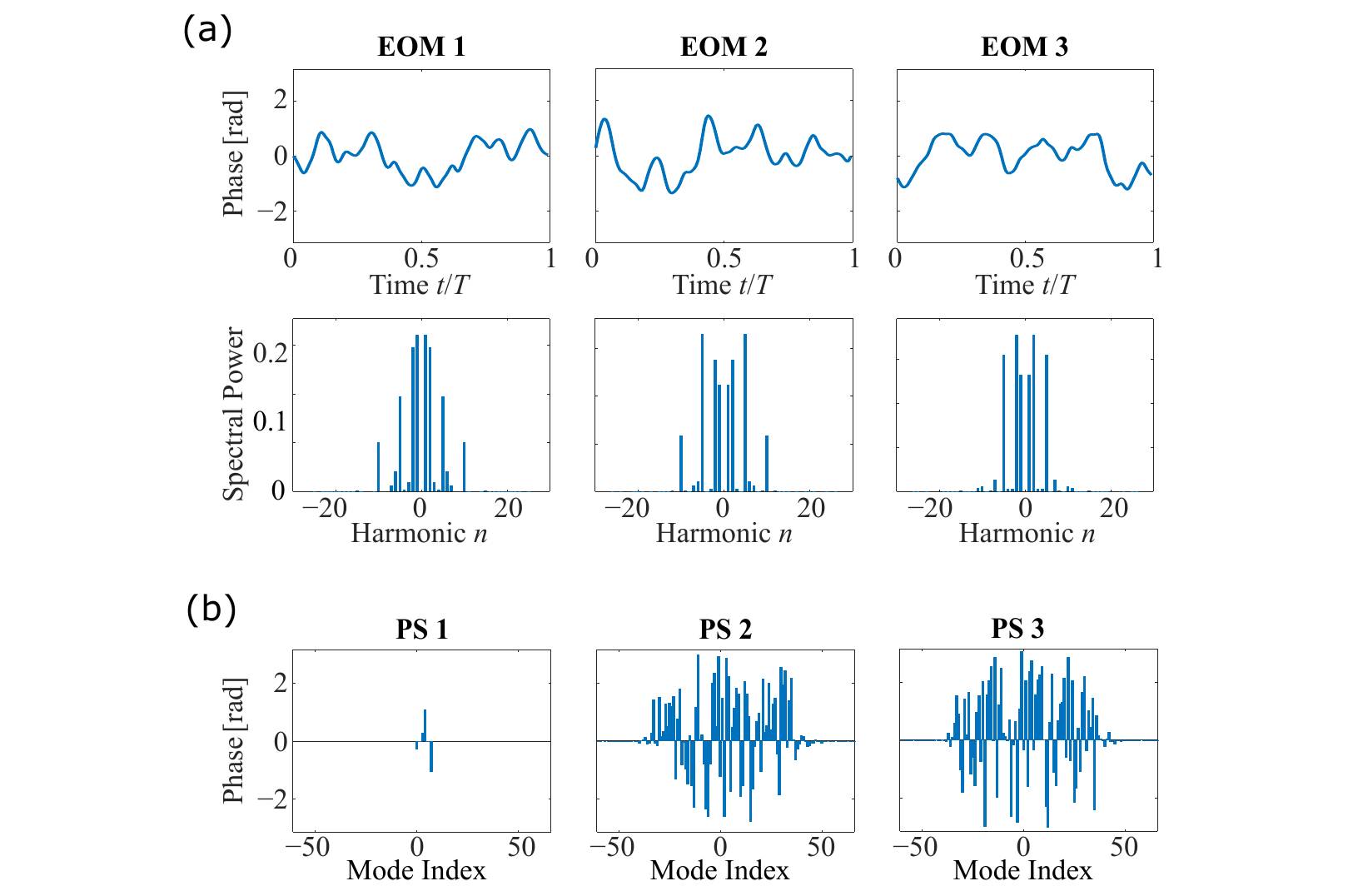}
\caption{\label{figS3} Phases applied to realize a spectral \textsc{cz} gate with $R=3$. Details for Fig. \ref{figS2} apply here as well.} 
\end{figure*}

\begin{figure*}
\includegraphics[width=6.5in]{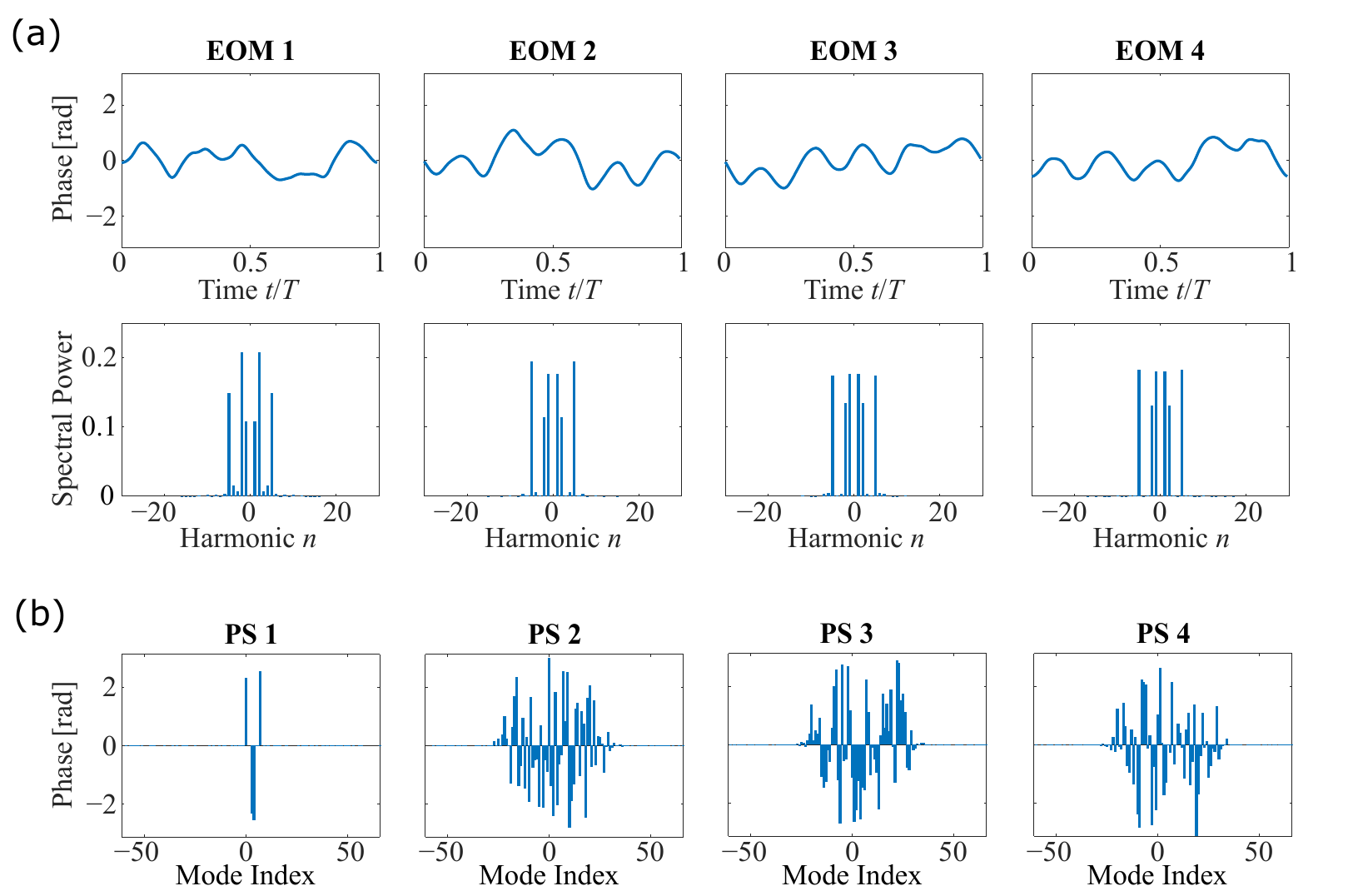}
\caption{\label{figS4}  Phases applied to realize a spectral \textsc{cz} gate with $R=4$. Details for Fig. \ref{figS2} apply here as well.} 
\end{figure*}

Finally, by mapping the indices $l$ and $m$ to their corresponding logical basis states, we establish the target transformation matrix $T$ for the particular operation considered, comparing it to $W$ via the fidelity $\mathcal{F}$, which we define according to the Hilbert-Schmidt inner product \cite{Uskov2009}:
\begin{equation}
\label{s9}
\mathcal{F} = \frac{\Tr(W^\dagger T) \Tr(T^\dagger W) }{\Tr(W^\dagger W) \Tr(T^\dagger T) }.
\end{equation}
The success probability $\mathcal{P}$ then follows as
\begin{equation}
\label{s10}
\mathcal{P} = \frac{\Tr(W^\dagger W) }{ \Tr(T^\dagger T) },
\end{equation}
a quantity which is meaningful (state-independent) whenever $\mathcal{F}=1$. And so, in the optimization procedure, we first define the specific mode mappings associating single-photon states with logical values [Eqs. (\ref{s3}), (\ref{s4}) or (\ref{s6}), (\ref{s7})], then use these to determine the desired state transformation $T$ for the targeted operation. Thereafter we run an optimization algorithm over the $2RM$ phases in Eq. (\ref{s1}) and propagated through to the matrix $W$, such that $\mathcal{F} = 1$ and $\mathcal{P}$ is maximized.

\section{Full Solutions}
\label{app2}
We enlist the Optimization Toolbox in MATLAB for our numerical routines. Our thorough---though certainly nonexhaustive---searches yield the results summarized in Table 1 of the main text. For completeness, here we record the specific phases for each pulse shaper and EOM in the optimal Hadamard and \textsc{cz} gates. All simulations truncate the space at $M=128$ modes, sufficiently precise without producing intractably long computation times.

Figure \ref{figS1} shows the results for the Hadamard gate. The specific temporal phases applied by each EOM are in Fig. \ref{figS1}(a), including the associated power spectra. These are not the mode coupling coefficients $c_n$ in Eqs. (6) and (7) of the article, but rather the Fourier transform of the temporal phase pattern $\varphi(t)$ itself, which indicates the electronic bandwidth required by an arbitrary waveform generator. The pulse shaper phases follow in Fig. \ref{figS1}(b). Their relatively discontinuous form poses no problems for devices which resolve each mode individually.

For the \textsc{cz} gate, we start with the same number of devices as in the Hadamard solution, adding pulse-shaper/EOM pairs until we reach the theoretical efficiency limit. Practically, this amounts to simulating the cases $R=2,3,4$ in Eq. (\ref{s1}). The specific EOM and pulse shaper phases for the $R=2$ case are presented in Fig. \ref{figS2}; those for $R=3$, in Fig. \ref{figS3}; and those for $R=4$, in Fig. \ref{figS4}. Note in particular how the EOM modulation bandwidths for these solutions decrease as the total number of devices increases: while the 10-dB bandwidth of the modulation patterns in Fig. \ref{figS2}(a) extend out to the 14th harmonic, the widest in Fig. \ref{figS3}(a) reaches just the 10th harmonic, and those of Fig. \ref{figS4}(a) stop at the 5th harmonic. This behavior corroborates the scaling described in the article, whereby solutions with more components are found to require fewer optical modes.


\begin{thebibliography}{10}
\bibitem{Note}
This electronic bandwidth is significantly smaller than the optical passband widths of Fig. \ref{fig3}, due to the nonlinear relationship between the phase in $\varphi(t)$ and the coefficients $c_k$ in Eq. (\ref{e5}), as well as the fact that cascaded EOMs excite more optical modes than possible by their bandwidths individually. See Appendix \ref{app2} for the specific microwave spectra.
\end{thebibliography}
\end{document}